\documentclass[aps,prl,reprint, superscriptaddress,nofootinbib]{revtex4-2}
\usepackage{graphicx}
\usepackage{dcolumn}
\usepackage{color}
\usepackage{bm}
\usepackage{amsmath, siunitx}
\usepackage{ulem}
\usepackage{gensymb}
\usepackage{float}
\usepackage{siunitx}

\newcommand{\basp}{BaFe$_2$(As$_{1-x}$P$_x$)$_2$}
\begin{document}

\title{Strange Metal Hall Effect in Underdoped BaFe$_2$(As$_{1-x}$P$_x$)$_2$}

\author{Augusto Ghiotto}
\affiliation{Department of Physics, University of California, Berkeley, California 94720, USA}
\affiliation{Materials Science Division, Lawrence Berkeley National Laboratory, Berkeley, California 94720, USA}

\author{Darian Hall}
\affiliation{Department of Physics, University of California, Berkeley, California 94720, USA}
\affiliation{Materials Science Division, Lawrence Berkeley National Laboratory, Berkeley, California 94720, USA}

\author{Yuanqi Lyu}
\affiliation{Department of Physics, University of California, Berkeley, California 94720, USA}
\affiliation{Materials Science Division, Lawrence Berkeley National Laboratory, Berkeley, California 94720, USA}

\author{Kohtaro Yamakawa}
\affiliation{Department of Physics, University of California, Berkeley, California 94720, USA}
\affiliation{Materials Science Division, Lawrence Berkeley National Laboratory, Berkeley, California 94720, USA}

\author{Sophie Rodehutskors}
\affiliation{Max Planck Institute for the Structure and Dynamics of Matter, Hamburg, Germany}

\author{Corina Dunn}
\affiliation{Department of Physics, University of California, Berkeley, California 94720, USA}
\affiliation{Materials Science Division, Lawrence Berkeley National Laboratory, Berkeley, California 94720, USA}

\author{Philip J. W. Moll}
\affiliation{Max Planck Institute for the Structure and Dynamics of Matter, Hamburg, Germany}

\author{John Singleton}
\affiliation{National High Magnetic Field Laboratory, Los Alamos National Laboratory, Los Alamos, NM, USA}

\author{Nikola Maksimovic}
\affiliation{Department of Physics, Boston University, Boston, MA 02215, USA}

\author{James G. Analytis}
\altaffiliation{Contact for correspondence, analytis@berkeley.edu}
\affiliation{Department of Physics, University of California, Berkeley, California 94720, USA}
\affiliation{Materials Science Division, Lawrence Berkeley National Laboratory, Berkeley, California 94720, USA}
\affiliation{Kavli Energy NanoScience Institute at the University of California, Berkeley
and the Lawrence Berkeley National Laboratory, Berkeley, California 94720, USA}


\begin{abstract}

The unusual transport properties of strange metals point to the breakdown of the quasiparticle picture, the understanding of which remains one of the most vexing problems in physics. Here, we report investigations of the electrical Hall effect of the strange metal superconductor BaFe$_2$(As$_{1-x}$P$_x$)$_2$. We show that a doping-independent contribution to the Hall effect exists within a fan shaped region above a putative quantum critical point. This `strange metal Hall' contribution echoes many of the properties of the antiferromangetic Hall response, but attains universal properties that distinguish it from the effects of Fermi surface reconstruction. This is consistent with an underlying origin connected to the presence of critical fluctuations, tying it to observations of $T$-linear resistivity and the appearance of unconventional superconductivity.

\end{abstract}

\maketitle


The nature of strange metals (SM) is among the most important open problems in condensed matter physics. They defy the existing paradigm of Fermi liquids in a solid, exhibiting linear-in-temperature dependence of resistivity down to very low temperatures with a scattering rate at the so-called Planckian scale \cite{Varma2020, Phillips2022}. Strange metals are prevalent across a multitude of distinct superconducting materials including the cuprates \cite{Taillefer2010}, iron-based superconductors \cite{Shibauchi2014}, heavy fermions \cite{Gegenwart2008} and moiré superlattices \cite{Cao2020, Ghiotto2021}, suggesting that a universal framework could underlie all of these systems. It has been suggested that this is tied to the presence of a quantum critical point (QCP) \cite{Paschen2020} where there could appear unconventional, fractionalized excitations, a possibility which continues to drive experimental studies \cite{Chien1991, Anderson1991, Senthil2008, Maksimovic2022}.

The transport behavior of these systems shows notable deviations from conventional Fermi liquids. Experiments in magnetoresistance have revealed the violation of Kohler's rule and a linear-in-field dependence of resistivity that scales with its temperature \cite{Harris1995, Analytis2014, Hayes2016, GiraldoGallo2018}. The Hall coefficient, $R_{\rm H}$, also displays unusual behavior, evolving roughly as $1/T$ which leads to the famous $T^2$ dependence of the cotangent of the Hall angle characterizing a possible `hidden Fermi liquid' \cite{casey_hidden_2011}. Recently, we identified that this low-temperature enhancement of the Hall coefficient in \basp\, can be suppressed by magnetic field leading to a field-dependent decay in $R_{\rm H}$. The decaying component appears to be additive in the Hall coefficient, which we identify as the strange metal Hall effect (SMHE). Importantly, similar effects are observed in many other classes of strange metals \cite{Chien1991b, Hayes2020}. In this work, we explore the connections of this enhancement to the antiferromagnetic state of $\text{BaFe}_2(\text{As}_{1-x}\text{P}_x)_2$ and argue that the SMHE emerges from an antiferromagnetic critical point that delocalizes carriers.


\begin{figure*}[t]
\includegraphics[width=\linewidth]{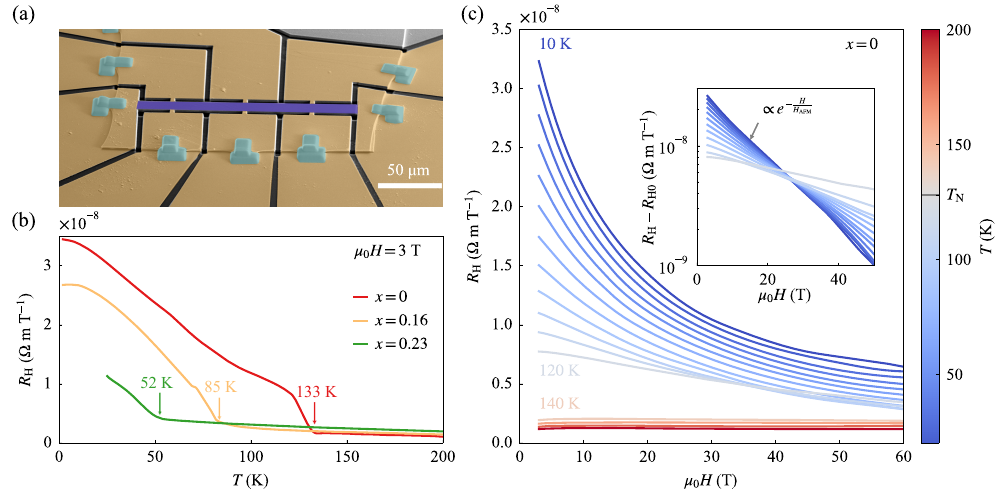}
\caption{\label{fig:fig1} \textbf{Hall effect in the underdoped BaFe$_{2}$(As$_{1-x}$P$_{x}$)$_{2}$}.
(a) Microstructured Hall bar of $\mathrm{BaFe}_2(\mathrm{As}_{1-x}\mathrm{P}_x)_2$. False-colored SEM image indicating the Hall bar (purple), gold contacts (gold) and platinum contacts (cyan).
(b) Temperature dependence of the Hall coefficient measured at 3 T for $x=0 ~ (T_{\rm N} \approx 133 K)$, $0.16  ~ (T_{\rm N} \approx 85 K, T_c \approx 2 K)$ and $0.23 ~  (T_{\rm N} \approx 52 K, T_c \approx 22 K)$. The Hall coefficient increases dramatically below $T_{\rm N}$ and displays a pronounced composition dependence.
(c) Magnetic field dependence of the Hall coefficient for $x=0$ at different temperatures (10 K to 120 K in steps of 10 K, 140 K to 200 K in steps of 20 K). The AFM state has a very clear exponential fall-off below $T_{\rm N}$ that is suppressed with increasing temperature (inset). Above $T_{\rm N}$, $R_{\rm H}$ rapidly saturates in field.
}
\label{FIG:intro}
\end{figure*}

The iron-based superconductor $\text{BaFe}_2(\text{As}_{1-x}\text{P}_x)_2$ is an ideal platform to study quantum criticality and the phenomenology of strange metals at accessible magnetic fields \cite{Kasahara2010,Shibauchi2014, Analytis2014, Hayes2016, Maksimovic2020, Hayes2020}. The parent compound exhibits an antiferromagnetic-nematic-orthorhombic order (henceforth AFM, for short), which is suppressed upon phosphorus substitution of arsenic, leading to a QCP at $x\approx0.31$ where the superconducting temperature is maximized and so is the slope of the $T-$linear resistivity. In this work, we report that $R_{\rm H}$ exhibits a fast fall-off with magnetic field in the AFM phase, superficially similar to the SMHE for the optimally doped $\text{BaFe}_2(\text{As}_{1-x}\text{P}_x)_2$, but an order of magnitude larger \cite{Hayes2020}. However, the fall-off is strongly composition dependent and its suppression tracks the suppression of the AFM order parameter with P-substitution, being maximized in the $x=0$ parent compound. The SMHE of the paramagnetic state on the other hand, remains roughly composition independent across the phase diagram, straddling the QCP in a fan-like shape. It abruptly vanishes for $x < 0.16$ where the Hall effect can be simply connected to band structure effects. Crucially, this cut-off coincides in doping with the edge of the superconducting dome.


Underdoped $\text{BaFe}_2(\text{As}_{1-x}\text{P}_x)_2$ crystals were grown using a $\text{Ba}_2\text{As}_3$:$\text{Ba}_2\text{P}_3$ flux method described elsewhere \cite{Nakajima2012}. In order to obtain sufficiently thin crystals for pulsed high field measurements, the crystals were cleaved with thermal release tape and transferred onto $\text{SiO}_2$/intrinsic-Si substrates, followed by the sputtering of $\approx 200$~nm of Au. Uniform flakes with thicknesses between 1 and 5~$\mu$m were identified using a Scanning Electron Microscope (SEM). Platinum (Pt) contacts bridging the substrate and the crystal were deposited via ion-assisted deposition. To minimize surface damage, the Au layer was selectively removed from the Hall bar region using an 8~kV $\text{Ga}^{+}$ Focused Ion Beam (FIB). The crystal was subsequently microstructured into a Hall bar geometry (as shown in Fig. 1(a)) with a 30~kV, $\approx 1$~nA $\text{Ga}^{+}$ FIB. Low-field Hall measurements were performed using conventional 4-terminal lock-in techniques. Pulsed high-field measurements were conducted at the Los Alamos National Laboratory using a custom setup. The applied current density was maintained on the order of $10^3$~A/cm$^2$. All of the presented Hall data have been antisymmetrized with respect to the magnetic field.


\begin{figure*}[t]
\includegraphics[width=\linewidth]{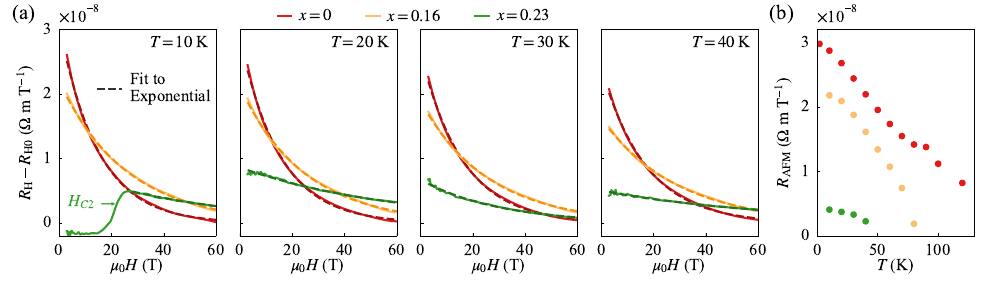}
\caption{\label{fig:fig2} \textbf{Composition dependence of the Hall coefficient in the AFM state.}  
(a) Hall coefficient at 10 K, 20 K, 30 K and 40 K for x = 0, 0.16 and 0.23. The enhanced Hall coefficient is present even when the zero-field ground state is superconducting. 
(b) Pre-factor $R_{AFM}$ for the exponential fall-off as a function of temperature and composition. Increasing temperature and P-doping suppress the low field enhancement of the Hall coefficient.
}
\label{FIG:UD_RH}
\end{figure*}



We begin by examining the Hall coefficient in the AFM state, defined as $R_{\rm H} = \rho_{xy}/\mu_0H$. The AFM $R_{\rm H}$ is highly composition dependent, with no indications of the presence of fluctuation phenomena. We show in Fig. \ref{FIG:intro}(b) the temperature dependence of the low-field Hall coefficient for $x = 0$, $0.16$, and $0.23$. $R_{\rm H}$ undergoes a pronounced enhancement as the system enters the reconstructed AFM phase, which grows as the temperature decreases and is  suppressed as P-substitution increases. $R_{\rm H}$ exhibits a sharp exponential-like magnetic field dependence in the AFM state (shown for $x = 0$ in Fig. \ref{FIG:intro}(c)) which is suppressed at elevated temperatures until it vanishes at $T_{\rm N}$. In the paramagnetic state ($T>T_{\rm N}$), $R_{\rm H}$ rises and saturates in field, which indicates that the system is imperfectly compensated with majority holes \cite{pippard_magnetoresistance_1989}.



In Fig. \ref{FIG:UD_RH} we show the evolution of the field-dependent $R_{\rm H}$ with $x$ at different temperatures below $T_{\rm N}$. Isovalent P substitution strongly suppresses the low-field Hall coefficient. The magnetic-field behavior bears striking similarity to the exponential form of the SMHE \cite{Hayes2020} in the paramagnetic state, so we parametrize it with the same phenomenological expression $R_{\rm H} = R_{AFM(SM)}e^{-H/H_0} + R_{\rm H0}$. In this parameterization, $R_{AFM(SM)}$ is an additive enhancement of $R_{\rm H}$ from zero to infinite field, $H_0$ allows us to stretch the exponential {\it ad-hoc}, and $R_{\rm H0}$ is the high field saturation value. The fit results for parameter $R_{AFM}$ as a function of temperature are shown in Fig. \ref{FIG:UD_RH}(b) for $x = 0$, $0.16$, and $0.23$ (the fits obtained for these curves have $R^2>0.98$). In essence, the contribution to the Hall effect inside the AFM is profoundly composition and temperature dependent, qualitatively tracking the suppression of the order parameter. In stark contrast with this behavior, the SMHE in the paramagnetic region  is roughly independent of P-substitution. This universal behavior spans overdoped and underdoped regions in a fan-like shape about the putative QCP at $x\approx0.31$. 

\begin{figure*}[t]
\includegraphics[width=\linewidth]{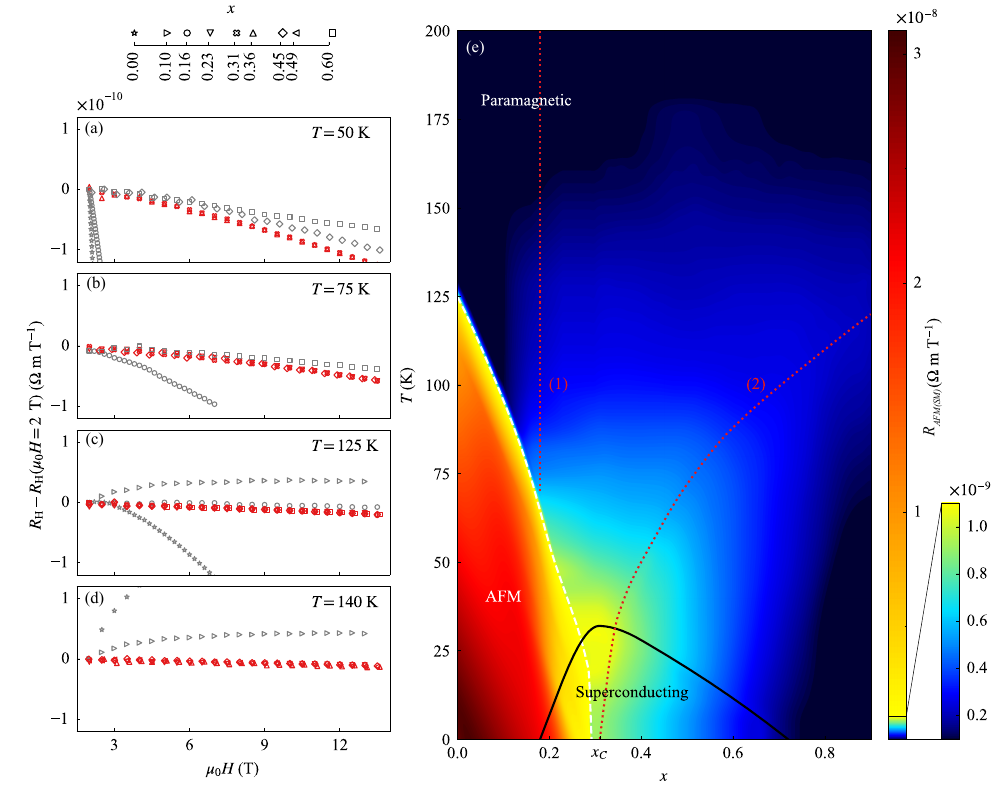}
\caption{\label{fig:fig3} \textbf{Universal behavior in the quantum critical fan.}
(a) At 50 K, the Hall fall-off for $0.31$ and $0.36$ collapses. For $0.16  ~ (T_{\rm N} \approx 85 K, T_c \approx 2 K)$, $R_{\rm H}$ falls off significantly faster as the system is in the AFM state. 
(b) At 75 K, the Hall fall-off for $0.23  ~ (T_{\rm N} \approx 52 K, T_c \approx 22 K)$ falls within the critical behavior of higher compositions. For $0.16  ~ (T_{\rm N} \approx 85 K, T_c \approx 2 K)$, $R_{\rm H}$ falls off faster as expected for the AFM state. 
(c) At 125 K, the quantum critical fall-off extends to x = 0.16, which is at the edge of the superconducting dome. For $x=0 ~ (T_{\rm N} \approx 133 K)$, $R_{\rm H}$ still displays a sharp fall-off in the AFM state. Surprisingly, for $x=0.10  ~ (T_{\rm N} \approx 100 K)$ the Hall coefficient rises before saturating at higher fields. 
(d) At 140 K, $R_{\rm H}$ increases before saturating for both $x=0 ~ (T_{\rm N} \approx 133 K)$ and $x=0.10  ~ (T_{\rm N} \approx 100 K)$, but still manifests a fall-off for $0.23 ~  (T_{\rm N} \approx 52 K, T_c \approx 22 K)$.
(e) Phase diagram of the exponential enhancement pre-factor $R_{AFM(SM)}$ for the AFM state and the quantum critical fan. In the AFM, $R_{AFM}$ is largest near x = 0 and at lower temperatures. In contrast, in the normal state $R_{SM}$ is enhanced near the quantum critical point, albeit an order of magnitude smaller than that of the AFM. Furthermore, $R_{SM}$ in the normal state is weakly temperature and composition dependent, being cut off away from the compositional range of superconductivity (lines (1) and (2)). All the data shown for $x>0.3$ is from ref. \cite{Hayes2020}. 
}
\label{FIG:collapse}
\end{figure*}

In Fig. \ref{FIG:collapse}(a-d), we plot the underdoped and overdoped regimes taken at different temperatures and subtract the low field value to highlight the field-dependent behavior of $R_{\rm H}$. In the quantum critical fan, the field dependence collapses for all compositions $x$ (red), defining the region of the SMHE. In the overdoped region, this behavior becomes non-universal below the critical fan, and all curves vary in field above the SMHE curves. In the underdoped region, the non-universal behavior appears within the AFM boundary, where curves always fall far below the SMHE, or in the conventional paramagnetic regions, where familiar behavior of a slightly uncompensated metal is recovered with a dependence always above the SMHE.


We summarize the phase diagram for the phenomenological pre-factor $R_{AFM(SM)}$, corresponding to data within the AFM (strange metal) phase, in Fig. \ref{FIG:collapse}(e). The range of quantum critical behavior is remarkably consistent with the quantum critical fan seen via $T-$linear resistivity \cite{Kasahara2010}.



\begin{figure}[t]
\includegraphics[width=\columnwidth]{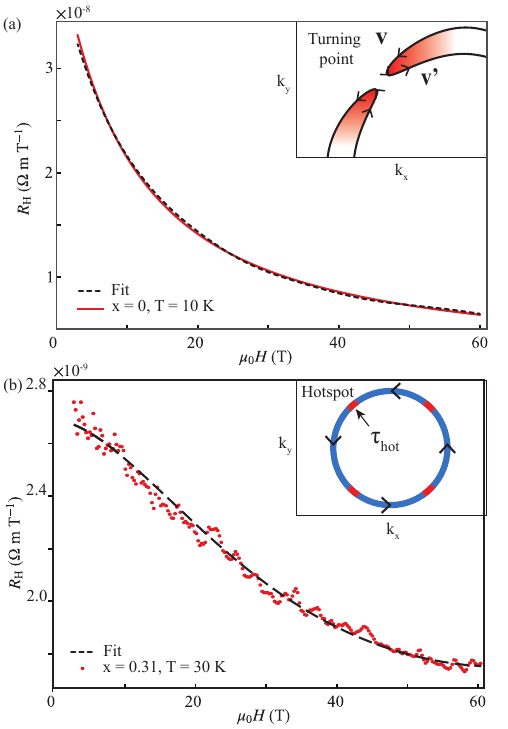}
\caption{\label{fig:fig4} \textbf{Fits to the turning point and hotspot models.}
(a) Turning point model fit to $R_{\rm H}$ of the parent compound assuming $\tau = 1.2 \times 10^{-13}$ s, an AFM gap of 62 meV and effective mass of 1.6$m_e$. The resulting curve has been renormalized in amplitude. \textit{Inset}: Turning point model of the imperfectly nested Fermi surface in the AFM phase. Turning points are regions of the Fermi surface where the charge velocity changes dramatically. 
(b) Hotspot model fit to $R_{\rm H}$ of the optimally doped compound assuming $\tau_{\rm holes} = 8.3 \times 10^{-14}$ s, $\mu_{\rm h}/\mu_{\rm e} = 1.2$ with a scattering anisotropy factor of 4.8. \textit{Inset}: Hotspot model of the normal state Fermi surface. Quantum fluctuations induce specific regions on the Fermi surface where the scattering lifetime is much shorter than in the rest of the Fermi surface. Data from ref. \cite{Hayes2020}.
}
\label{FIG:cartoon}
\end{figure}

To understand the properties of Fig. \ref{FIG:collapse}, first we consider how the AFM reconstruction of the Fermi surface affects the Hall coefficient. The reconstruction introduces sharp features known as turning points (see Fig. \ref{FIG:cartoon}(a)), whose contribution can be calculated using the Shockley-Chambers integral \cite{Koshelev2013, Maksimovic2020}. It has been shown that such a model effectively captures the magnetotransport properties of \basp \cite{Koshelev2013, Maksimovic2020}, and so we build a toy model along these lines (see Fig. \ref{FIG:cartoon}(a) and the S.I.). Disorder in this calculation plays an important role because it emphasizes areas of the FS with the greatest curvature, so that the turning points dominate   $R_{\rm H}$ \cite{Ishida}. As the magnetic field increases, the Fermi velocity at the turning points is efficiently averaged, and $R_H$ approaches values comparable to the unreconstructed FS even before the AFM gap is closed. The calculated $R_{\rm H}$ falls off with increasing magnetic field in agreement with the data (shown in Fig. \ref{FIG:cartoon}(a)). 


When we extend this model to the strange metal regime, which exhibits qualitatively similar $R_{\rm H}$ vs $H$, we encounter a shortcoming that suggests strong fluctuations may underlie the SMHE. In a similar simplified model of turning points, we calculate the Hall coefficient considering anisotropic scattering rates, known as ``hotspots", acting on the normal state bands (see Fig. 4(b) and the  S.I.). In these hotspots, the scattering lifetime is orders of magnitude smaller than in the ``cold" regions of the Fermi surface, where transport is dominated. Such a model is a natural extension of the effects of fluctuations of the order parameter onto the normal state bands, and has been originally proposed to explain the Fermi arcs observed in the pseudogap of the cuprates \cite{Abanov} and has been widely applied to explain the magnetotransport in a variety of systems that exhibit linear-in-field magnetoresistance  \cite{Koshelev2016, Maksimovic2020, Hinlopen2022, Kool2026}. Within a range of realistic parameters, we can get good agreement with the data; a declining field-dependent $R_{\rm H}$ which is similar in form but an order of magnitude smaller than what is seen in the AFM (Fig. \ref{FIG:cartoon}(b)). This is not surprising, since the connection between hotspots and turning points is very natural given the former arise when in proximity to an AFM transition \cite{rosch_magnetotransport_2000}.

There is an important contrast between the field-dependent fall-off for the AFM and the paramagnetic regions of the phase diagram. For fixed temperature, the field-dependent fall-off depends strongly on $x$ in the AFM, diminishing quickly upon approaching the phase boundary. In the paramagnetic state the SMHE is negligible until $x \approx 0.16$, and instead the Hall coefficient behaves as expected in an imperfectly compensated Fermi liquid (first rising in field and then saturating in both $x=0$ and $x=0.10$ shown in Fig. \ref{FIG:collapse}(c, d)). But at $x=0.16$, $R_{SM}$ onsets abruptly and is nearly independent of $x$. This is compatible with the edge of the quantum critical fan demarcating the boundary of the SM, and perfectly consistent with the onset of linear-in-$T$ resistivity seen in ref. \cite{Kasahara2010}; there is no phase transition upon entering the SM, but transport properties attain a universal character. This low field behavior is unique to the SMHE phenomenology.

This observation betrays an irksome tension between our model and the data: reproducing the $x$ dependence of $R_{\rm H}$ in the strange metal state requires a fine-tuned balance between electron and hole mobilities across the phase diagram --- an unlikely scenario given that P-substitution alters the disorder level in the material in a way that affects the hole and electron carriers very differently \cite{Shishido2010}. Changes in disorder can skew the fall-off of the Hall coefficient as evidenced in our calculations, where we observe large changes in the field-dependent $R_{\rm H}$ for a modest ($\approx $50\%) difference in relative mobilities (see S.I.). We note that electrons are known to be much more mobile than holes in these compounds, differing by up to a factor of 10 in some cases \cite{Shishido2010}, emphasizing that our model does not naturally capture this important feature. The additive nature of the SMHE and the indifference to disorder are therefore highly suggestive of a scale-invariant mechanism.

A magnetic field suppression of the Hall coefficient has also been reported in other strange metal systems such as LSCO \cite{Shekhter}, FeSe$_{1-x}$S$_{x}$ \cite{ulo2021}, cuprates \cite{Putzke2021}, and essentially in all systems where $T-$linear resistivity is observed with a $T^2$ cotangent of the Hall angle. However, $\text{BaFe}_2(\text{As}_{1-x}\text{P}_x)_2$ is a unique platform to explore this phenomenon because the isovalent substitution simplifies the analysis by maintaining a compensated carrier density across the phase diagram, and the metallic nature of the AFM state allows us to track the transverse transport continuously from the strange metal down into the symmetry-broken phase. In the AFM state, the exponentially decreasing field dependence of $R_{\rm H}$ arises because the magnetic field begins to overwhelm the reconstruction of the FS, effectively pacifying the turning points which dominate at low fields, giving way to larger parts of the FS which dominate at high fields. The hotspot picture tells us that nearly AFM quasiparticles behave in an analogous manner, crossing over from scattering dominated by small regions of the FS to large regions of the FS. However, this explanation is incomplete because it cannot explain the insensitivity of the SMHE to the P content $x$. This independence points to a more fundamental picture of where quasiparticles near an AFM-to-paramagnetic delocalization transition lead to universal transport signatures.

Finally, we note that the underdoped edge of the critical fan in Fig. \ref{FIG:collapse}(e) coincides with the edge of the superconducting dome, which mirrors what occurs in the overdoped regime \cite{Hayes2020}. This striking feature of the \basp\ phase diagram highlights that seemingly competing phases, magnetism and unconventional superconductivity, share intimate ties: the present data connects the SMHE to the AFM state, and its ``shadow" indicates the presence of a scattering mechanism that ultimately leads to unconventional superconductivity.

\section{Acknowledgments}
This work was supported by the Brown Foundation. A.G. thanks the Miller Institute for support. Work at the Molecular Foundry was supported by the Office of Science, Office of Basic Energy Sciences, of the U.S. Department of Energy under Contract No. DE-AC02-05CH11231, under user proposal number MFP-10837. We acknowledge support from the US National Science Foundation (NSF) Grant Number 2201516 under the AccelNet program of the Office of International Science and Engineering (OISE). A portion of this work was performed at the National High Magnetic Field Laboratory (NHMFL), which is supported by National Science Foundation Cooperative Agreement Nos. DMR-1644779 and DMR-2128556 and the Department of Energy (DOE). J.S. acknowledges support from the DOE BES program “Science at 100 T”.

\bibliographystyle{apsrev4-2} 
\bibliography{references}    

\end{document}